\documentclass{optica-article}

\journal{opticajournal} 

\articletype{Research Article}

\usepackage{lineno}
\usepackage{graphicx}
\usepackage{dcolumn}
\usepackage{bm}
\usepackage{textcomp}
\usepackage{siunitx}

\begin{document}

\title{Heterogeneous tantala photonic integrated circuits for sub-micron wavelength applications}

\author{Nima~Nader,\authormark{1,*} Eric~J.~Stanton,\authormark{1,3,4} Grant~M.~Brodnik,\authormark{2,3} Nusrat~Jahan,\authormark{1,3} Skyler~C.~Weight, \authormark{1,3} Lindell~M.~Williams,\authormark{2,3} Ali~Eshaghian~Dorche,\authormark{1,3,5} Kevin~L.~Silverman,\authormark{1} Sae~Woo~Nam,\authormark{1} Scott~B.~Papp,\authormark{2} and Richard~P.~Mirin,\authormark{1}}

\address{\authormark{1}Applied Physics Division, National Institute of Standards and Technology, 325 Broadway, Boulder, CO 80305, USA\\
\authormark{2}Time and Frequency Division, National Institute of Standards and Technology, 325 Broadway, Boulder, CO 80305, USA\\
\authormark{3}Department of Physics, University of Colorado, 2000 Colorado Avenue, Boulder, CO 80309, USA\\
\authormark{4}Currently with EMode Photonix, Boulder, CO 80305, USA\\
\authormark{5}Currently with Nexus Photonics, 6500 Hollister Avenue, Suite 150, Goleta, CA 93117, USA}

\email{\authormark{*}nima.nader@nist.gov} 


\begin{abstract*}
Atomic and trapped-ion systems are the backbone of a new generation of quantum-based positioning, navigation, and timing (PNT) technologies. The miniaturization of such quantum systems offers tremendous technological advantages, especially the reduction of system size, weight, and power consumption. Yet, this has been limited by the absence of compact, standalone photonic integrated circuits (PICs) at the wavelengths suitable for these instruments. Mobilizing such photonic systems requires development of fully integrated, on-chip, active components at sub-micrometer wavelengths. We demonstrate heterogeneous photonic integrated circuits operating at 980~\si{nm} based on wafer-scale bonding of InGaAs quantum well active regions to tantalum pentoxide passive components. This high-yield process provides $>$~95~\% surface area yield and enables integration of $>$~1300 active components on a 76.2 mm (3 inch) silicon wafer. We present a diverse set of functions, including semiconductor optical amplifiers, Fabry-Perot lasers, and distributed feedback lasers with 43~dB side-mode suppression ratio and $>$~250~GHz single-mode tuning range. We test the precise wavelength control and system level functionality of the on-chip lasers by pumping optical parametric oscillation processes in microring resonators fabricated on the same platform, generating short-wavelength signals at 778~\si{nm} and 752~\si{nm}. These results provide a pathway to realize fully functional integrated photonic engines for operation of compact quantum sensors based on atomic and trapped-ion systems.
\end{abstract*}

\section{Introduction}
Quantum systems based on optically-controlled neutral atoms and trapped ions are a critical component of several emerging technologies. These include quantum information processing with trapped-ion qubits \cite{Ion:Computing, PRXQuantum.2.020343}, optical atomic clocks \cite{OpticalClock:Italy, RevModPhys.87.637, McGrew:19}, cold-atom interferometers \cite{PhysRevLett.97.010402, PhysRevApplied.12.014019, Lee2022-wu} for Global Positioning System (GPS) free navigation, magnetometers \cite{Budker2007-mn, Shah2007-xe}, and gravimeters \cite{Rosi2014-hc}. Given the broad application base, a significant effort has been placed on developing such systems with reduced form factor through microfabrication of small-scale vapor cells \cite{microCell:Kitching:APL, microCell:Hummon:OL} and ion traps \cite{PhysRevLett.96.253003}, and integration with photonic circuits \cite{Hummon:18, Newman:19, Niffenegger2020-nk, Mehta2020-wr, Isichenko2023-ng, doi:10.1126/sciadv.ade4454}. Despite this progress, quantum systems haven't reached their full scalability and miniaturization potential. This is partly due to the limited access to on-chip laser sources and heterogeneous photonic integrated circuits (PICs) at the required wavelengths for the control of such systems, including  visible and ultraviolet wavelengths \cite{OpticalClock:Italy, RevModPhys.87.637, Gallagher_1994, Holloway:Nat:Rev, Inlek:PhD, Imreh:PhD, brownnutt:PhD, Ball:ionTrap}.

The demand from datacom and telecom industries has resulted in a rapid development of heterogeneous PICs at wavelengths around 1310 nm and 1550 nm \cite{Tin:IEEEreview, XIANG202335, Bowers:IEEETutorial, Shekhar2024-hs}. These platforms are based on integration of InAs quantum dot  \cite{Shang2021-fi} and InP-based quantum well (QW) \cite{Fang:06} gain regions with silicon-on-insulator photonics. Such platforms do not support sub-micrometer wavelengths, due to the type of gain media and the silicon bandgap absorption. The inability to use silicon photonics at these shorter wavelengths necessitates development of platforms involving large bandgap, low refractive index, photonic materials based on deposited dielectrics. Recently, short-wavelength heterogeneous PICs have been demonstrated based on chip-scale bonding of GaAs-based actives on silicon nitride (Si\textsubscript{3}N\textsubscript{4}) with system functionality at 980 nm \cite{Nexus:Nat2022} and 780 nm \cite{Nexus:Optica:2023}.

In this paper, we present a new integrated photonic platform based on heterogeneous integration of InGaAs-on-GaAs QW active regions and tantalum pentoxide (Ta\textsubscript{2}O\textsubscript{5}, also known as tantala) passive regions. Tantala is an ultra-low loss, deposited dielectric with a wide transparency window spanning from the ultraviolet to the mid-infrared wavelengths and a refractive index of $\approx$~2.11 at 980~\si{nm} \cite{Black:OL21}. Tantala offers key material advantages over Si\textsubscript{3}N\textsubscript{4} that make it an appealing photonic platform for integration with active components. Namely, it can be deposited at room temperature using ion beam sputtering, requires a much lower annealing temperature of $<$~600~\textdegree{}C to achieve low material absorption, has a lower residual stress, and has a smaller thermo-optic coefficient (\si{\num{8.8e-6}}~\si{1/K}). This III-V semiconductor/tantala platform offers sub-micrometer functionality at 980~\si{nm} with various components such as semiconductor optical amplifiers (SOAs), Fabry-Perot (FP) lasers, and tunable, single-mode distributed feedback (DFB) lasers. We further show the utility of the single-mode lasers by pumping wide-span degenerate optical parametric oscillation (OPO) processes based on the third order, $\chi^{(3)}$, nonlinearity \cite{Kartik:OPO:2019} in dispersion-engineered microring resonators.

\section{Wafer-scale fabrication}
Heterogeneous PICs consist of III-V epitaxial layer stacks integrated with passive photonics on a common substrate. Fabrication processes developed for such integration are predominantly based on chiplet-level integration methods such as micro-transfer printing \cite{Ghent} and chip-scale bonding \cite{Nexus:Nat2022}. In micro-transfer printing, the active devices are formed on their native substrate (InP or GaAs) and then released and transferred to the Si-based photonic circuit using a polydimethylsiloxane (PDMS) stamp \cite{Ghent}. This process simplifies the processing by enabling fabrication of active components on their native substrate but requires the design of misalignment-tolerant structures due to its coarse alignment accuracy of $\pm$~500~\si{nm} \cite{Ghent, 10.1063/1.5120004}. In the case of chip-scale bonding, it enables co-fabrication of passive photonics and active components with precise alignment limited by the lithography step. This process also enables bonding of III-V materials with different designs on the same photonic wafer. The final device yield is, however, limited by the number of bonded III-V chips and their surface area \cite{Tran:24}.

\begin{figure}[ht!]
\centering\includegraphics[width=10cm]{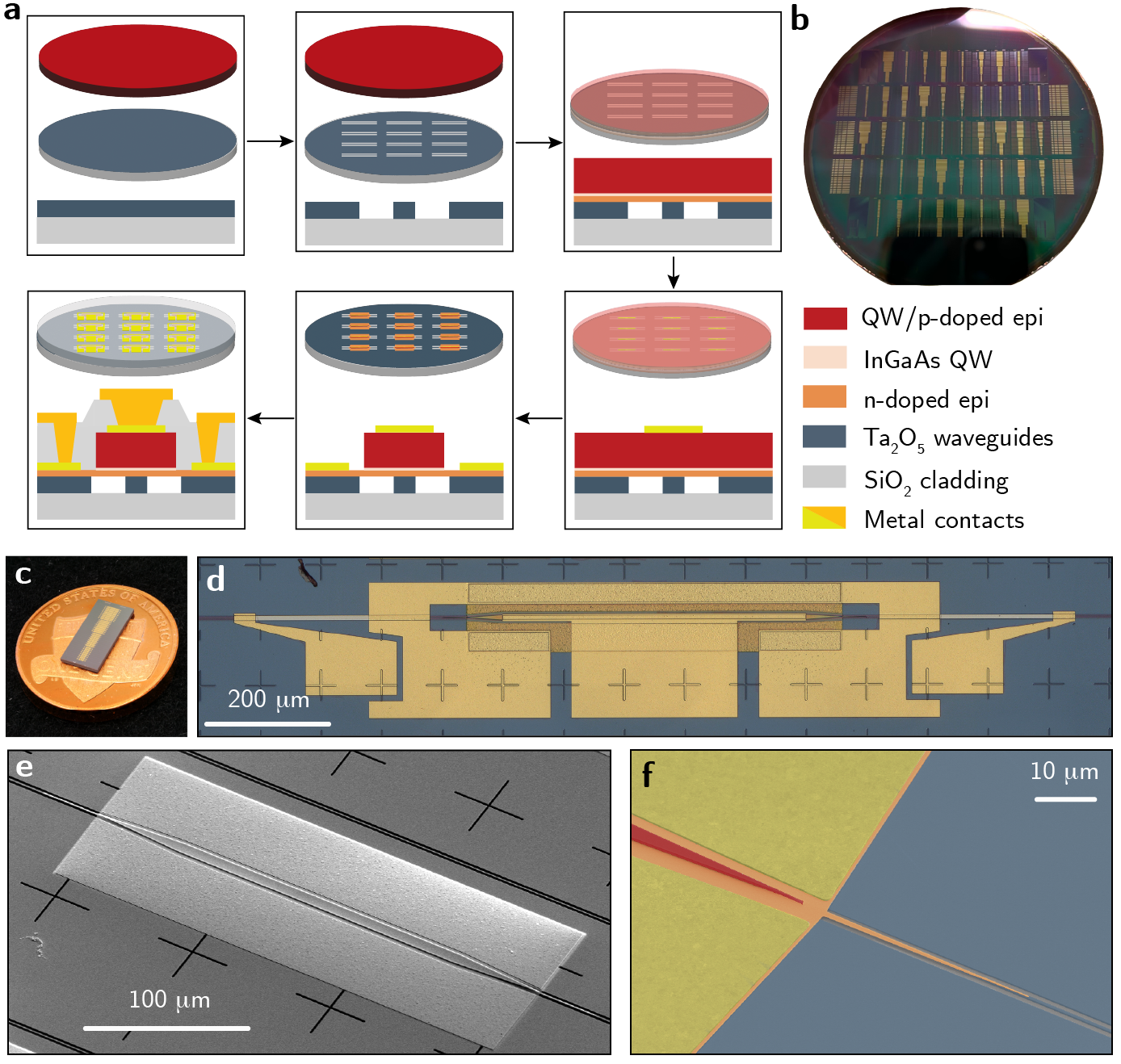}
\caption{\label{fab}Wafer-scale fabrication of heterogeneous tantala PICs. (a) Simplified process flow depicting the wafer scale bonding of the III-V material stack and formation of the active components. (b) Camera image of the fabricated 76.2 mm wafer after SiO\textsubscript{2} top-cladding deposition. (c) Focus-stacked camera image of a fully functional chip with 32 integrated active components. (d) Microscope image of an on-chip DFB laser. (e) Scanning electron micrograph (SEM) of a patterned and etched III-V active region integrated on tantala waveguides. (f) Zoomed-in, false color, SEM of the III-V to tantala transition taper structure. Here, the colors match the color scheme in (a).}
\end{figure}

We develop a fabrication process based on 76.2 mm (3 inch) wafer-scale bonding of a III-V semiconductor layer stack to a silicon wafer with tantala PICs. The wafer-scale bonding process enables utilization of the entire wafer, with $>$ 95 $\%$ surface area yield, to achieve dense integration of active and passive components with $<$ 10 nm alignment accuracy (limited by lithography instrumentation). We utilize this process to yield $>$ 40 chips with more than a thousand integrated active components. Figure~\ref{fab}a presents the simplified wafer-scale process flow. We grow the AlGaAs/InGaAs active layers lattice-matched to a 76.2 mm GaAs substrate using molecular beam epitaxy. The epitaxial layer stack consists of two 7.25 nm thick In\textsubscript{21}Ga\textsubscript{79}As QW layers separated by 10 nm thick GaAs barriers \cite{Dorche:OFC, Dorche:CLEO} (see supplement 1 for the full epitaxial layer stack). A 570 nm thick tantala film is ion-beam sputtered on an oxidized Si wafer with 800 nm thick thermal SiO\textsubscript{2} to form the waveguiding layer of the PIC. The waveguide structures, including the routing waveguides, chip-couplers, and feedback structures required for operation of integrated lasers, are formed by electron beam lithography (EBL) and dry etch processes based on CHF\textsubscript{3}/CF\textsubscript{4}/Ar plasma. The 76.2 mm GaAs epi-wafer is then directly bonded to the tantala wafer after using atomic layer deposition (ALD) to deposit 12 nm of an Al\textsubscript{2}O\textsubscript{3} interlayer and 
O\textsubscript{2} plasma surface activation. After bonding, the GaAs substrate is removed with a NH\textsubscript{4}OH/H\textsubscript{2}O\textsubscript{2} wet etch process to expose the epitaxial semiconductor layer stack (refer to Fig. 1s of supplement 1 for a camera image of the wafer-scale bonded epitaxy on the tantala wafer).  

To form the electrically-pumped active components (Fig.~\ref{fab}e), we first form contacts to the \textit{p}-type Be-doped GaAs layer on top of the III-V layer stack through lift-off of a deposited Ti/Au/Ti (5 nm/50 nm/5 nm) film. Next, the laser mesa and the bottom \textit{n}-contact structures are EBL patterned and dry-etched by a BCl\textsubscript{3}/Cl\textsubscript{2}/Ar chemistry. Following the III-V etch, 10 nm ALD Al\textsubscript{2}O\textsubscript{3} is deposited to passivate the etched sidewalls of the III-V structure, and electrical contact to the \textit{n}-type Si-doped semiconductor is formed by lift-off of a 50 nm/100 nm/250 nm/5 nm thick Pd/Ge/Au/Ti metal stack. We then deposit a 1.4 \textmu{}m thick layer of $\mathrm{SiO_{2}}$ as optical top-cladding for the tantala waveguides and electrical isolation for the diodes. Figure~\ref{fab}b presents a camera picture of the processed 76.2 mm wafer after the SiO\textsubscript{2} top-cladding deposition. Fabrication of the lasers is completed by via etching into the top-cladding to access the contact metals and electron beam deposition of 1 \textmu{}m thick gold probe pads (Fig.~\ref{fab}c-d).

A crucial component enabling heterogeneous PICs is a coupling structure that facilitates efficient light coupling from the III-V material to passive waveguides. Achieving an efficient mode transition from the high refractive index III-V device ($n\textsubscript{eff}\approx3.32$) to the low index tantala waveguide ($n\textsubscript{eff}\approx1.95$) requires design of complex photonic structures. One possible design is based on the use of an intermediary material as a mode converter \cite{Nexus:Nat2022}. In our design, the coupling structure is based on multistage inverse tapers etched at the two ends of the III-V active region (Fig.~\ref{fab}f). The high refractive index of the III-V results in an optical mode that is mostly localized in the III-V mesa. We first taper the III-V ridge from the primary active region width to a 100 nm tip (red colored taper in Fig.~\ref{fab}f). This reduces the effective modal index such that it enables efficient optical mode transition to the underlying \textit{n}-contact layer. The \textit{n}-contact layer (orange colored area in Fig.~\ref{fab}f) is then tapered to a 100 nm tip in two stages to facilitate light coupling to the tantala waveguide with a simulated 86 $\%$ efficiency (the details of the multistage taper design are provided in supplement 1).

\section{Integrated single-mode lasers}
Single-mode lasers will be at the heart of integrated photonic engines designed for the control and operation of compact atomic systems \cite{Nexus:Nat2022, Nexus:Optica:2023, Srinivasan:NatPhot:2023}. Distributed feedback (DFB) lasers are one of the most widely used single-mode lasers due to their simple cavity design, high side-mode suppression ratio (SMSR), and long-term stability. A free running DFB laser, however, exhibits a typical linewidth of a few megahertz, which is broader than required for many applications in quantum science and technology. Recently, it has been demonstrated that the frequency noise of these lasers can be suppressed by $>$ 20 dB \cite{SILDFB:Vahala:2023, SILDFB:Bowers:Science2021} to outperform stabilized fiber laser sources \cite{Bowers:FiberLaser:OL2021} when they are self-injection locked to high quality factor, low mode volume microresonators in low-loss PICs. 

\begin{figure}[ht!]
\centering\includegraphics[width=11cm]{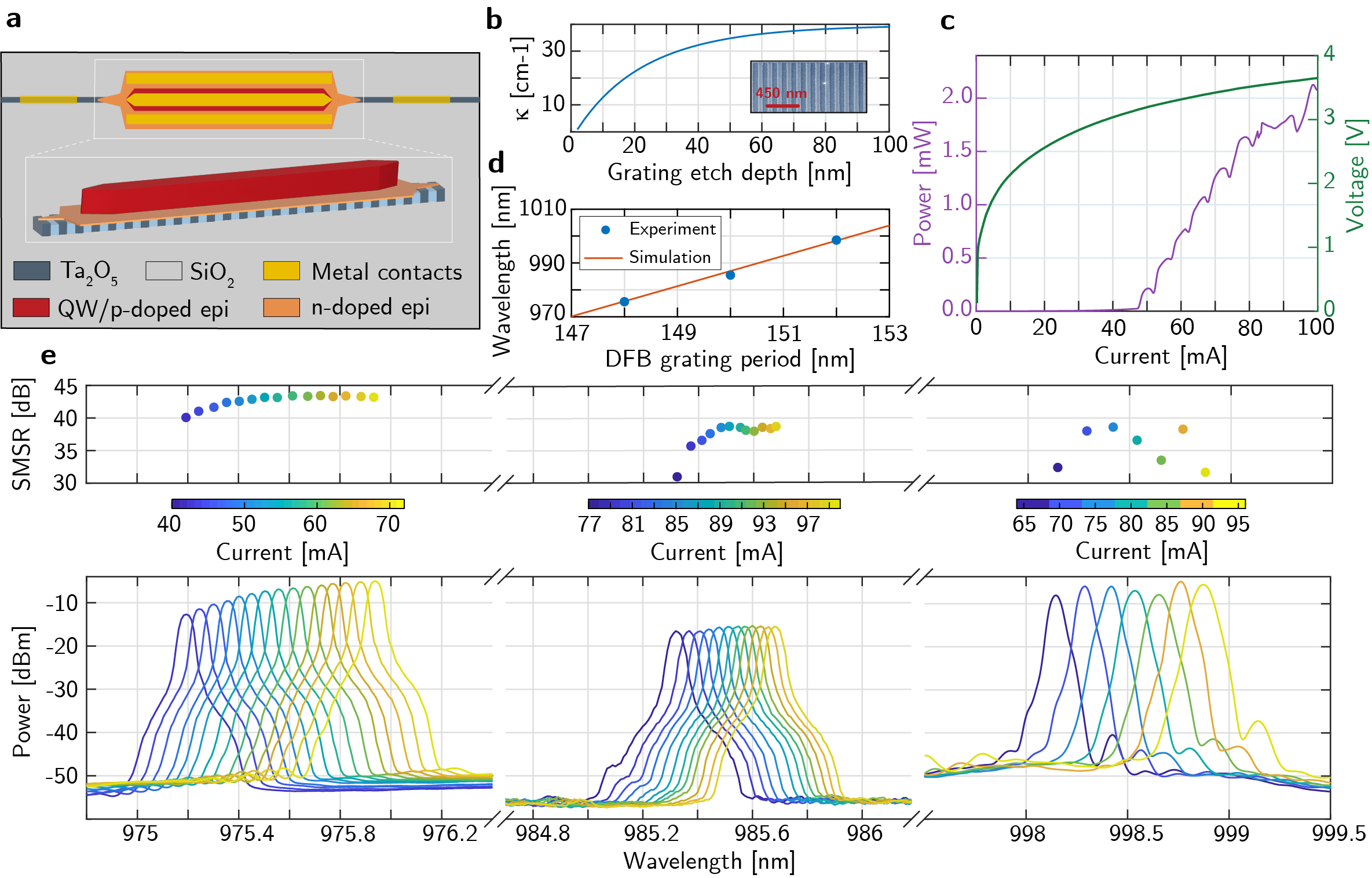}
\caption{\label{LIV}Optical testing of the DFB lasers. (a) Schematic diagram of the DFB cavity with the low-$\kappa$ feedback gratings etch in the tantala waveguides. (b) Calculated DFB grating strength, $\kappa$, as a function of the tantala etch depth. (c) Measured laser power-current and voltage-current curves of a fabricated device operating at 998 nm. (d) Calculated and measured lasing wavelengths as a function of DFB grating period showing excellent agreement within 1.5~\si{nm}. (e) Measured spectra and SMSR of lasers operating at 975 nm, 985 nm, and 998 nm, in left, center, and right panels, respectively.} 
\end{figure}

We fabricate and characterize DFB lasers on the III-V/tantala platform based on the cavity design in Ref.~\cite{Fang:DFB:OE2008, Fang:DFB:OE2011}. Figure~\ref{LIV}a presents a schematic diagram of the DFB cavity. In the active area, we form the DFB grating as corrugations in the height of the tantala waveguide (Fig.~\ref{LIV}b, inset). This takes advantage of the small optical power overlap with the tantala waveguide structure ($\mathrm{\approx~0.5~\%}$) to achieve a low grating feedback strength.  A quarter-wave phase shift element is placed in the middle of the grating structure to facilitate a single lasing mode at the Bragg frequency of the grating \cite{Coldren}. Figure~\ref{LIV}b presents the calculated etch-depth-dependent feedback strength as $\kappa=(1/\Lambda)(\Delta n/\overline{n})$ with $\Lambda=(\lambda\textsubscript{0}/4)(1/n\textsubscript{0}+1/n\textsubscript{1})$, $\Delta n~=~|n\textsubscript{1}-n\textsubscript{0}|$, and $\overline{n}=(n\textsubscript{0}+n\textsubscript{1})/2$ \cite{Coldren}. Here, $\lambda\textsubscript{0}$ is the feedback wavelength, and $n\textsubscript{1}$ and $n\textsubscript{0}$ are the effective modal indices of the grating segments with and without corrugation, respectively. We target grating etch depth of 8~\si{nm}, corresponding to feedback strength of $\kappa$~=~10~\si{cm^{-1}} in a 1.2~\si{mm} long active area, resulting in cavities with $\kappa L~=~1.2$ and effective mirror reflectances of 29~$\%$. 

We mount the laser chips on a temperature-controlled stage for optical testing and characterization at stage temperatures ranging from 6 \textdegree{}C to 35 \textdegree{}C. We sweep the injection current through the diode while monitoring the voltage drop across the device and the output optical power of the laser. To monitor the output power, we butt-couple a wide-area photodiode to one facet of the chip and collect the generated light through 12~degree angled facet couplers. The measured laser voltage (green) and optical power (purple) curves for a DFB laser with grating period of 152 nm are plotted in Fig.~\ref{LIV}c as a function of the injection current. The data is measured at 20 \textdegree{}C stage temperature. The measured threshold current is 45 mA with maximum recorded output power of 2 mW at 100 mA injection current. The kinks in the output power are attributed to heat-related longitudinal mode-hoping in the laser cavity \cite{Bowers:EDBR:Optica2020}. The lasing mode is defined by the spectral alignment of the DFB grating resonance to one of the longitudinal modes of the cavity at a wavelength with high spectral gain. As the injection current is increased, the III-V gain spectra redshifts at a higher rate than the DFB grating resonance. This results in the detuning of the cavity modes from the peak grating feedback causing the drop in the output power until eventually the next longitudinal mode is aligned. 

To measure the optical spectrum, we couple the laser output to a single-mode 980 nm lensed fiber (3~dB~$\pm$~1~dB coupling efficiency) and record the spectrum with an optical spectrum analyzer (OSA) with a resolution bandwidth limit of 0.05 nm (15.6~GHz). Figure~\ref{LIV}e presents the recorded spectra and side-mode suppression ratio (SMSR) of three DFB lasers with grating periods of 148 nm, 150 nm, and 152 nm  on the left, center, and right panels, respectively. Phase heater elements on the two sides of the active region (Fig.~\ref{fab}d) enable a current-dependent, mode-hop free tuning range of ~$>$~250~\si{GHz} with tuning rates of 26.5~\si{pm/mA}, 16.6~\si{pm/mA}, and 24.0~\si{pm/mA}, respectively (refer to supplement 1). Figure~\ref{LIV}d presents the calculated and measured lasing wavelengths of the DFB lasers as a function of the feedback grating period. We observe excellent agreement between the designed and measured wavelengths within 1.5~\si{nm}. The lasing wavelength is calculated using $\lambda~=~2\overline{n}\Lambda$. For accurate calculation of $\overline{n}$, we model the heat transfer in the semiconductor diode using COMSOL multiphysics to simulate the QW device temperature at a certain injection current. We then estimate $n\textsubscript{0}$ and $n\textsubscript{1}$ at the simulated device temperature using the refractive index data for tantala \cite{Black:OL21} and the III-V material stack \cite{10.1063/1.373462}.

\begin{figure}[ht!]
\centering\includegraphics[width=6cm]{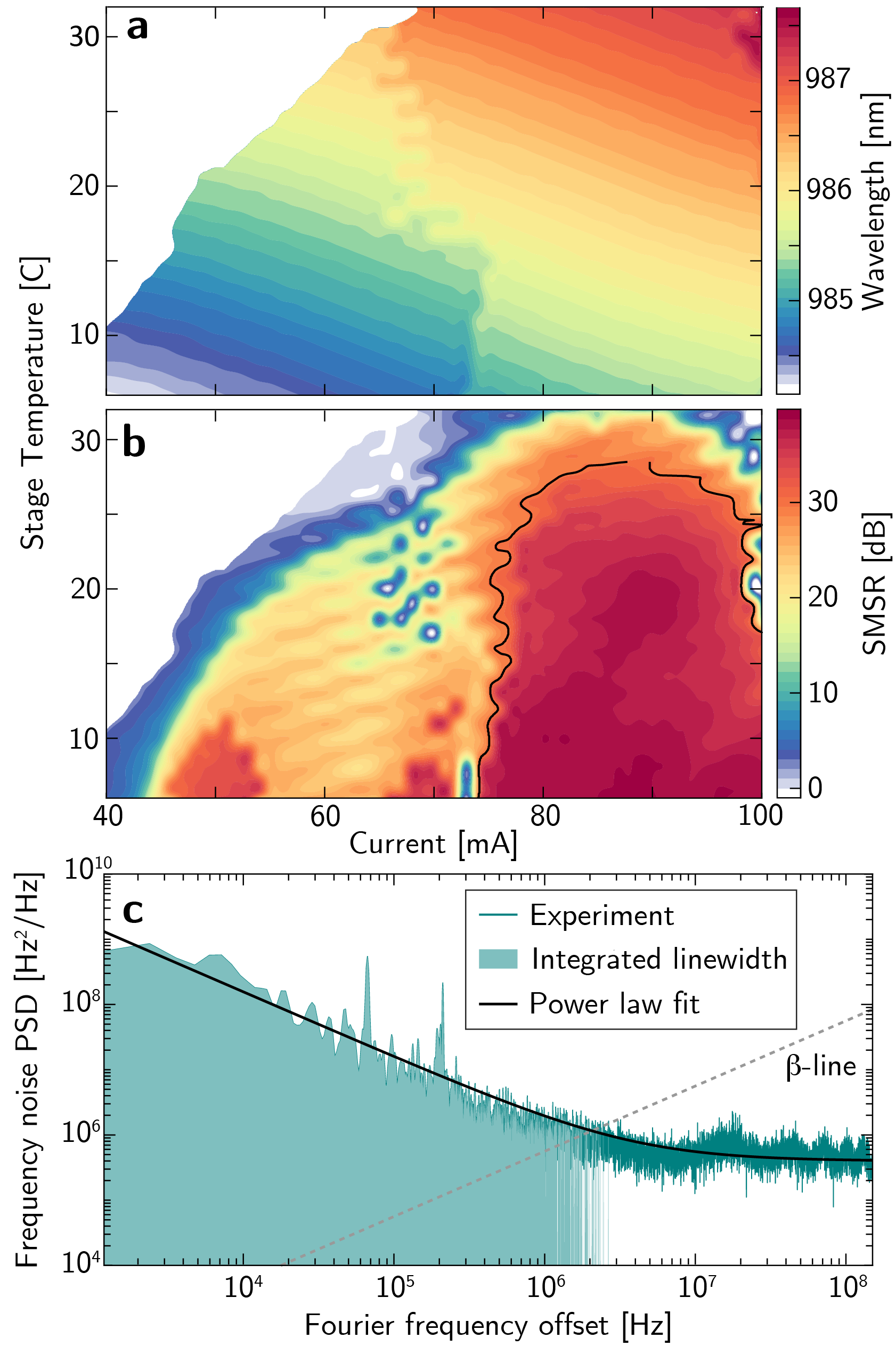}
\caption{\label{2D}Single-mode operation and frequency noise characterization of DFB lasers. Two dimensional mapping of the (a) lasing wavelength and (b) side-mode suppression ratio as a function of increasing stage temperature and injection current for a laser with feedback grating period of 150~\si{nm}. (c) Frequency noise power spectral density of a laser operating at 976~\si{nm} (feedback grating period 148~\si{nm}, injection current 76~mA).}
\end{figure}

To better understand the single-mode operation of the lasers, we plot a two-dimensional map of the DFB peak lasing wavelength and SMSR for a laser operating at 985~\si{nm} as a function of injection current (horizontal axis) and stage temperatures (vertical axis) in Figs.~\ref{2D}a and \ref{2D}b, respectively. To identify the single-mode operation region, we adjust the stage temperature from 6~\textdegree{}C to 32~\textdegree{}C in 1~\textdegree{}C increments, ensuring stability at each setpoint with a tolerance of ±5~m\textdegree{}C. At each temperature setpoint, the injection current is swept from 40~\si{mA} to 100~\si{mA} in 1~\si{mA} increments, and the optical spectrum is recorded using an OSA. The white areas in Figs.~\ref{2D}a and \ref{2D}b depict the current and temperature settings at which the laser is operating sub-threshold. A redshift in the peak lasing wavelength is visible in Fig.~\ref{2D}a as a function of increasing injection current and stage temperature. We observe a longitudinal mode-hop in the lasing wavelength at injection currents around 65-75~mA, depending on the stage temperature. At currents below this level, the DFB laser has a multi-mode behavior as evidenced by the low SMSR. After the mode-hopping, the laser operates in a single-mode region where the SMSR stays $>$~30 dB for a wide range of injection currents (70~\si{mA} to 100~\si{mA}) and stage temperatures (6~\textdegree{}C to 28~\textdegree{}C).

We measure the power spectral density (PSD) of the laser frequency noise (FN) using a fiber-based, calibrated, unbalanced Mach-Zehnder interferometer (MZI). The MZI has a free spectral range of 800~\si{MHz} and acts as an optical frequency discriminator. Figure~\ref{2D}c presents the measured FN PSD at the injection current of 76~\si{mA} for a DFB laser operating at 976~\si{nm} with 43 dB SMSR. At Fourier frequency offsets of $<$ 1~\si{MHz} the FN PSD scales inversely with the frequency as expected for a DFB laser. The FN PSD flattens at the white noise level of \si{\num{4e5}}~\si{Hz^2/Hz}, indicating a laser fundamental linewidth of 1.2~\si{MHz}, limited by the cavity length \cite{PhaseNoise}. We calculate the integrated linewidth of the laser based on two methods, namely the \si{1/\pi} \cite{1overPI} and the $\beta$-separation line \cite{DiDomenico:10}. In the \si{1/\pi} method, the integrated linewidth is defined as the Fourier frequency offset at which the accumulated phase noise, \textit{i.e.} the area under the FN PSD curve, reaches \si{1/\pi}~\si{rad^2}. For the data in Fig.~\ref{2D}c we calculate the accumulated phase noise starting from the highest frequency offset of the measurement (150~\si{MHz}), resulting in the integrated linewidth of 2.0~\si{MHz}. When applied to an ideal laser source with white frequency noise, the \si{1/\pi} method results in the full-width-half-maximum of the Lorentzian lineshape \cite{Liang:NatCom2015}. The $\beta$-line method gives a better estimate of the integrated linewidth for laser sources with added frequency noise other than the white noise since it also accounts for technical and flicker noise. In this method the frequency noise spectrum is divided in two regions separated by the $\beta$-separation line $S_{\delta\nu}~=~8\ln(2)f/\pi^{2}$ (dashed line in Fig.~\ref{2D}c). Here, $S_{\delta\nu}$ is the FN PSD and $f$ is the Fourier frequency offsets. In the FN PSD region with $S_{\delta\nu}$~$<$~$\beta$-line, the noise frequency is too fast and its modulation index is too slow to cause a significant broadening to the laser linewidth. Conversely, in the region with $S_{\delta\nu}$~$>$~$\beta$-line (highlighted with green in Fig.~\ref{2D}c), the noise level is high compared to the Fourier frequency and it causes a Gaussian broadening of the laser lineshape. In this method the integrated laser linewidth is calculated as the area under the FN PSD curve in the high modulation index region. We calculate the integrated linewidth of 9.4 MHz using the $\beta$-separation line method.

\section{System-level operation and outlook}
\begin{figure}[h!]
\centering\includegraphics[width=1.0\textwidth]{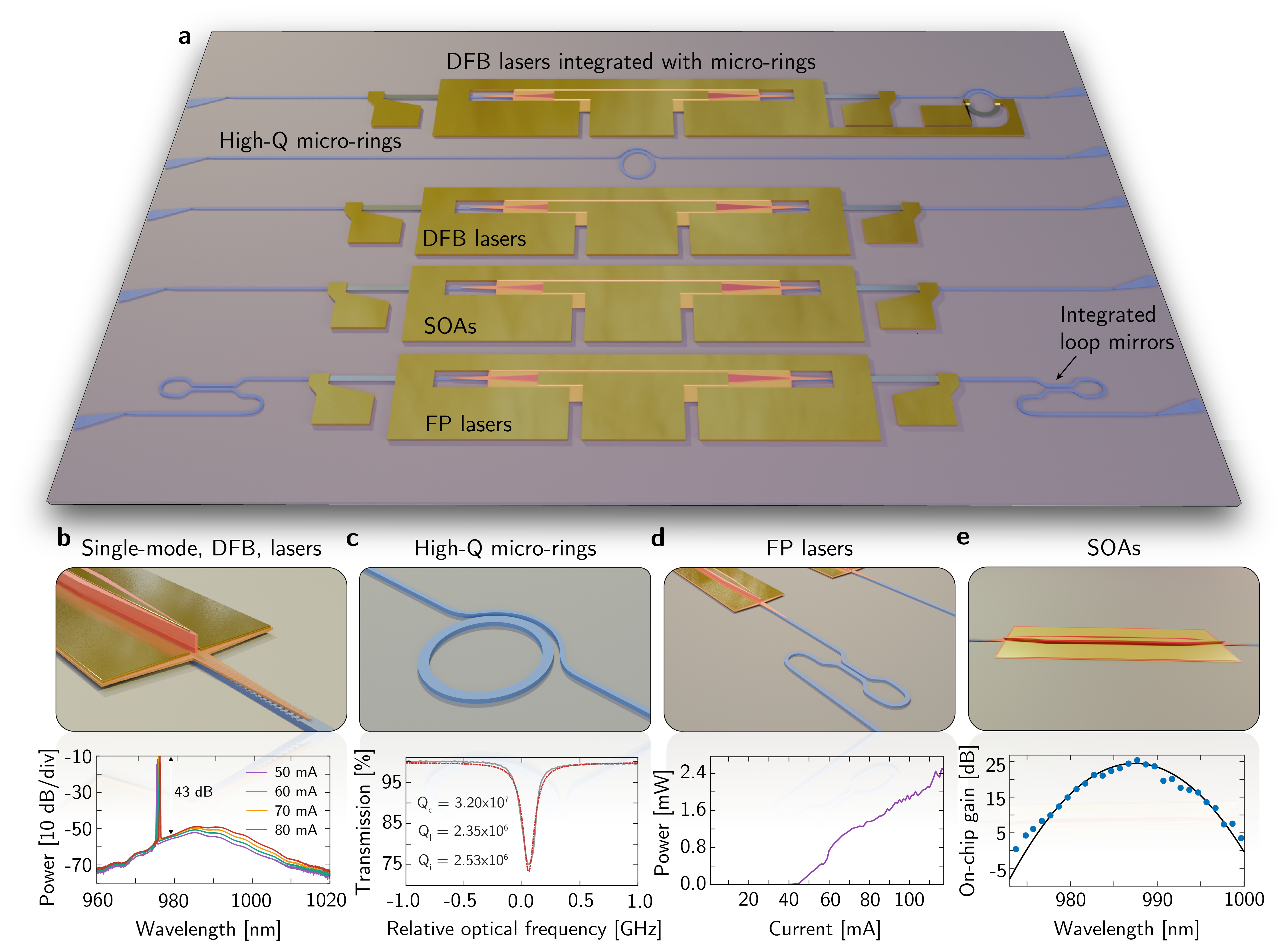}
\caption{\label{platform} (a) three-dimensional rendering of the fabricated heterogeneous tantala PIC supporting versatile, wafer-scale, integration of active components with low-loss passive PICs. (b-e) Active and passive functionalities demonstrated on the platform. From left to right: (b) Single-mode lasers based on distributed feedback cavities, (c) ultra low-loss passive tantala waveguides and high quality factor microring resonators, (d) Fabry-Perot (FP) lasers with integrated loop-mirrors etched in the tantala waveguides, and (e) tantala-integrated SOA with 24.5 dB small signal gain at 987.4~\si{nm}.}
\end{figure}

Figure~\ref{platform}a presents a three-dimensional schematic of a heterogeneous III-V/tantala PIC, highlighting the versatile integration of passive and active components with different functionalities. In addition to single-mode lasers (Fig.~\ref{platform}b), we demonstrate a variety of photonic components. The low optical loss in the deposited tantala layer, along with the SiO\textsubscript{2} top-cladding, enables fabrication of high quality factor (high-Q) microring resonators \cite{Hojoong:TaO:21}. These resonators are essential for applications requiring high-performance cavities, such as narrow-band filters and efficient nonlinear optical processes. Figure~\ref{platform}c presents a high-Q microring with intrinsic quality-factor of \num{2.5e6} at 1064~\si{nm}, fabricated with a ring width of 2.0~\textmu{}m. Figure~\ref{platform}d demonstrates a 1.5~\si{mm} long FP laser with 40~\si{mA} threshold current and $>$~2~\si{mW} waveguide-coupled output power. SOAs are an integral part of the photonic systems. Figure~\ref{platform}e presents the on-chip gain spectrum of a 2~\si{mm} long integrated SOA operating at 105~\si{mA} injection current for different wavelengths. Such devices are often used to enable system functionality in applications requiring high optical power such as nonlinear optical interactions for frequency comb generation and wavelength conversion in microring resonators. Here, the scattered data points are measured using a tunable laser coupled to the TE0 mode of the SOAs with on-chip input power of 25~\textmu{}W. The black curve represents the fit to the data using $G(\lambda)~=~G_p\exp{[-A(\lambda-\lambda_p)^2]}$, where $G_p$ is the peak on-chip gain, $\lambda_p$ is the peak gain wavelength, and $A$ is related to the 3-dB gain bandwidth by $\Delta\lambda~=~2\sqrt{\ln{2}/A}$ \cite{Volet:SOA}. We calculate the peak gain, wavelength, and 3-dB bandwidth of 24.5~dB, 987.4~\si{nm}, and 8.5~\si{nm}, respectively. We also calculate the peak modal gain, $g_{net}(\lambda_p)$, of 28.2~\si{\per\cm} from the $G_p$ factor using $G_p~=~\exp{[g_{net}(\lambda_p)L]}$ \cite{Volet:SOA}.

\begin{figure}[ht!]
\centering\includegraphics[width=6cm]{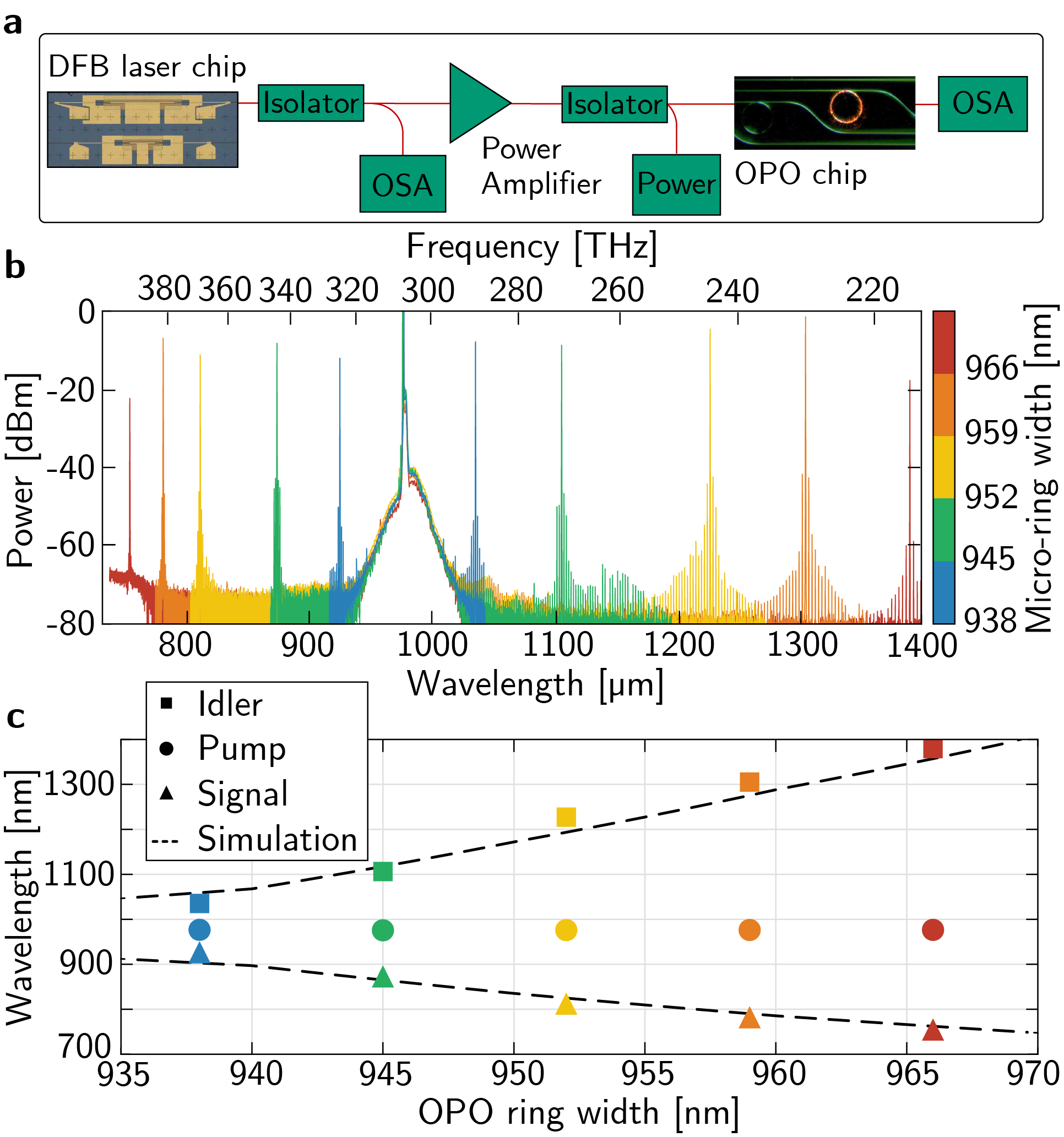}
\caption{\label{OPO} OPO pumped with the integrated DFB laser (a) DFB laser output is coupled into a single-mode fiber and amplified to 100~\si{mW} level before being coupled to a second tantala chip with the high quality factor (high-Q) rings. Fiber-coupled optical isolators are used to prevent back-reflections from entering the laser cavity. The output spectra of the high-Q resonators are monitored with an OSA. (b) Measured output OPO spectra of the microring resonators with different ring widths. (c) Simulated (dashed line) and measured (scattered dots) wavelengths of the generated signal and idler colors as a function of microring widths for the DFB pump wavelength of 976~\si{nm}. The data points are colored to match their corresponding OSA spectra in part (b).}
\end{figure}

Emerging quantum integrated photonic systems require on-chip lasers operating in the visible spectrum, extending down to ultraviolet wavelengths. Targeted wavelength conversion of a common pump laser through nonlinear optical interactions such as OPO \cite{Kartik:OPO:2019, Kartik:VISOPO:20, Black:OPO:22, Grant:OPO:arxiv:2024} and second harmonic generation \cite{Park:22} is shown to be a viable path for on-chip generation of short, hard-to-access wavelengths, specifically within the "green gap" \cite{Kartik:VISOPO:20, Kartik:OPO:GreegGap} and blue colors \cite{Park:22}. We test the utility of the integrated DFB lasers for such nonlinear optical interactions by pumping OPO processes in microring resonators fabricated in a separate tantala PIC. Fig.~\ref{OPO}a presents the experimental setup for OPO pumping. We couple the laser output (operating at 976~\si{nm}, Fig.~\ref{LIV}e, left panel) into a 980~\si{nm} single-mode fiber. We then use a fiber-coupled optical amplifier to amplify the laser light to $>$~100~\si{mW}, required to reach the OPO process threshold. The amplified light is then edge coupled into a separate tantala PIC with the fabricated microring resonators using lensed fibers with 6~dB coupling loss. For OPO pumping, we current- and temperature-tune the laser wavelength into the resonance with the TE0 optical mode of the microresonators and monitor the output optical spectra with an OSA (Fig.~\ref{OPO}b). 

The microrings of different widths provide phase matching conditions to generate signal and idler lights at different wavelengths. Figure~\ref{OPO}c compares the measured signal and idler wavelengths (scattered data points) with theoretical designs (dashed lines) for different ring widths. We design the OPO process in microring resonators by following the method outlined in \cite{Kartik:OPO:2019, Kartik:VISOPO:20} for geometrical group-velocity dispersion (GVD) engineering of the resonator waveguides. The resonator GVD can be controlled by tailoring the waveguide thickness and width, and the resonator radius. This enables control of the OPO phase matching condition through geometrical tuning of the effective modal index of the pump, signal, and idler waves. To emphasize platform compatibility for future heterogeneous integration of lasers and OPO resonators, we design microring geometries in a tantala layer with similar thickness used for the laser integration (570~\si{nm}). The further choice of resonator radius of 45~\textmu{}m and resonator waveguide widths in the range of 930~\si{nm}~-~970~\si{nm} facilitate normal GVD values at the pump wavelength of 976~\si{nm}, required to realize wide-span OPO processes \cite{Kartik:VISOPO:20}. Increasing the microring width results in generation of wider span OPO spectra and enables short-wavelength signal generation. As presented with orange and red curves in Fig.~\ref{OPO}b, we record generated signal wavelengths of $~$778~\si{nm} and 750~\si{nm} at the output of microrings with widths of 959~\si{nm} and 966~\si{nm}, respectively. The generated signal at 778~\si{nm} is suitable for integration with emerging portable atomic clocks based on two-photon absorption in Rb-87 atomic vapor cells \cite{Rb:2PA}.

The OPO pumping setup in Fig.~\ref{OPO}a includes two isolators to prevent back-reflections into the integrated DFB laser cavity. Heterogeneous photonic integrated circuits suffer from the lack of reliable on-chip isolators with optical power independent high isolation ratio ($>$~30~dB).  Alternatively, in an integrated laser+OPO system (depicted in Fig.~\ref{platform}a, top device), the back-reflections from the microring can be engineered for linewidth narrowing of the pump laser through self-injection locking \cite{Briles:SIL}. This would be possible by utilizing a phase heater element on the bus waveguide connecting the laser output to the microresonator. Such a heater element can be utilized to tune the phase of the back-reflected waves to achieve self-injection locking \cite{SILDFB:Bowers:Science2021} and enable isolator-free photonic circuits \cite{Xiang2023-ha}.

\section{Conclusion}
Ultra-low material loss and advanced fabrication processes have established tantala as a prime material platform for a variety of on-chip linear and nonlinear photonic applications including dielectric metasurface photonics \cite{Zhang2024-sp}, frequency comb generation \cite{Hojoong:TaO:21, 10.1063/5.0191602}, supercontinuum generation \cite{Lamee:20, Hojoong:TaO:21}, and wide-span OPO \cite{Black:OPO:22, Grant:OPO:arxiv:2024}. Here, we introduced active functionalities on tantala PICs through heterogeneous integration with InGaAs-on-GaAs QW gain material. The integration method introduced here enables high yield, wafer-scale, co-fabrication of passive and active components with $>$~95~\% surface area yield on a 76.2 mm wafer.  Integration of III-V gain sections with passive waveguides and external cavities enables demonstration of FP lasers, SOAs with $>$~21~dB gain, and current-tunable DFB lasers operating at 980~\si{nm} wavelength band with 43~dB side-mode suppression ratio and single-mode tuning range wider than 4$\times$ the longitudinal mode spacing of 60 GHz. 

We utilized our DFB laser as a pump source for $\chi^{(3)}$ microring resonator-based degenerate OPO to demonstrate the system-level functionality and utility of the on-chip active components. The resonators were designed and fabricated on the same 570~\si{nm} thick tantala platform used for the laser integration to show the potential for future integration of pump lasers diodes with mircoring resonators for monolithic OPOs. Our widest signal-to-idler frequency span is 183~\si{THz}, and the shortest wavelength signal generated is at 752~\si{nm}, limited by the available resonator geometries.

The heterogeneous integration presented in this work is agnostic to the design of the GaAs-based QW layer stack. We can take advantage of such versatility to realize photonic integrated circuits with active functionalities at shorter wavelengths down to $\approx$ 700~\si{nm}. Utilizing this platform, one can envision a heterogeneous PIC with active components operating in 700~\si{nm} to 980~\si{nm} wavelength range. In such a platform, pump lasers would be integrated with a variety of passive components designed for different nonlinear optical interactions such as frequency comb generation, and nonlinear wavelength conversion to shorter, hard to access, wavelengths. Fully functional photonic systems can be designed to support generation, routing, and manipulation of multiple wavelengths needed for operation of quantum systems based on specific atomic vapor cells or trapped ions. Such integrated quantum technologies have the potential to reshape a variety of high-impact industries through scalable miniaturization. As such, the deployment of these systems can mark a paradigm shift by offering robust, portable, and potentially autonomous (GPS-free) PNT systems.

\begin{backmatter}
\bmsection{Funding}
This work is supported by National Institute of Standards and Technology (NIST-on-a-chip program), the Defense Advanced Research Projects Agency (DARPA), Microsystems Technology Office (MTO) under the Lasers for Universal Microscale Optical Systems (LUMOS) program, AFOSR FA9550-20-1-0004 Project Number 19RT1019, and NSF Quantum Leap Challenge Institute Award OMA - 2016244.

\bmsection{Acknowledgment}
We thank Jeffrey Shainline, Scott Diddams, John Kitching, and David Carlson for useful discussions and inputs on the manuscript. Product disclaimer: Any mention of commercial products is for information only; it does not imply recommendation or endorsement by NIST. 

\bmsection{Disclosures}
The authors declare no competing interests. EJS is the co-founder at EMode Photonix.

\bmsection{Data Availability}
Data underlying the results presented in this paper are not publicly available at this time but may be obtained from the authors upon reasonable request.

\bmsection{Supplemental document}
See Supplement 1 for supporting content.
\end{backmatter}

\bibliography{Optica-template}
\end{document}


\maketitle

\section{Quantum well gain}

The 980~\si{nm} quantum-well gain region is based on AlGaAs/InGaAs active layers lattice-matched to a 76.2 mm diameter GaAs substrate using molecular beam epitaxy (MBE). Table \ref{epiStack} presents the MBE-grown layer stack. The gain region consists of two 7.25~\si{nm} thick In\textsubscript{0.21}Ga\textsubscript{79}As strained quantum wells (QWs) separated by 10~\si{nm} thick GaAs barriers. To achieve low series resistance in the laser diode, we choose Be- and Si-doped GaAs layers as p- and n-type contact layers, respectively. The highly doped n- and p-type contact layers are grown with \num{2e18}~\si{\per\cubic\cm} and \num{3e19}~\si{\per\cubic\cm} doping concentrations, respectively. A 800~\si{nm} thick Al\textsubscript{0.6}Ga\textsubscript{0.4}As layer is grown as the top-cladding to maximize optical mode confinement in the QW region. We also include a 50~\si{nm} thick graded-index separate-confinement-heterostructure (GRIN-SCH layer) to provide carrier confinement in the active region and minimize current leakage.

\begin{table}[htbp]
\centering
\caption{\label{epiStack}\bf Epitaxial layer stack for the quantum-well 980~\si{nm} active region. QW: Quantum-well, GRIN-SCH: graded-index separate-confinement-heterostructure, UID: undoped.}
\begin{tabular}{ m{1.5cm} m{3.5cm} m{2cm} m{2cm} m{2.25cm} }
\hline
Layer & Material & Thickness~[\si{nm}] & Doping~[\si{\per\cubic\cm}] & Refractive index \\
\hline
n-contact & GaAs & 150 & n-, \num{2e18} & 3.52 \\
Barrier & GaAs & 10 & UID & 3.52 \\
QW & In\textsubscript{0.21}Ga\textsubscript{0.79}As & 7.25 & UID & 3.84 \\
Barrier & GaAs & 10 & UID & 3.52 \\
QW & In\textsubscript{0.21}Ga\textsubscript{0.79}As & 7.25 & UID & 3.84 \\
Barrier & GaAs & 10 & UID & 3.52 \\
GRINSCH & Al\textsubscript{0.60}Ga\textsubscript{0.40}As --> GaAs & 100 & p-, \num{7e17} & 3.19 --> 3.52 \\
p-cladding & Al\textsubscript{0.60}Ga\textsubscript{0.40}As & 800 & p-, \num{7e17} & 3.19 \\
p-grading & GaAs --> Al\textsubscript{0.60}Ga\textsubscript{0.40}As & 800 & p-, \num{1e18} & 3.52 --> 3.19 \\
Barrier Reduction & GaAs & 50 & p-, \num{3e18} & 3.52 \\
p-contact & GaAs & `00 & p-, \num{3e19} & 3.52 \\
Etch-stop & Al\textsubscript{0.80}Ga\textsubscript{0.20}As & 150 & UID &  3.10 \\
\hline
\end{tabular}
  \label{tab:shape-functions}
\end{table}

\begin{figure}[htbp]
\centering
\includegraphics[width=10cm]{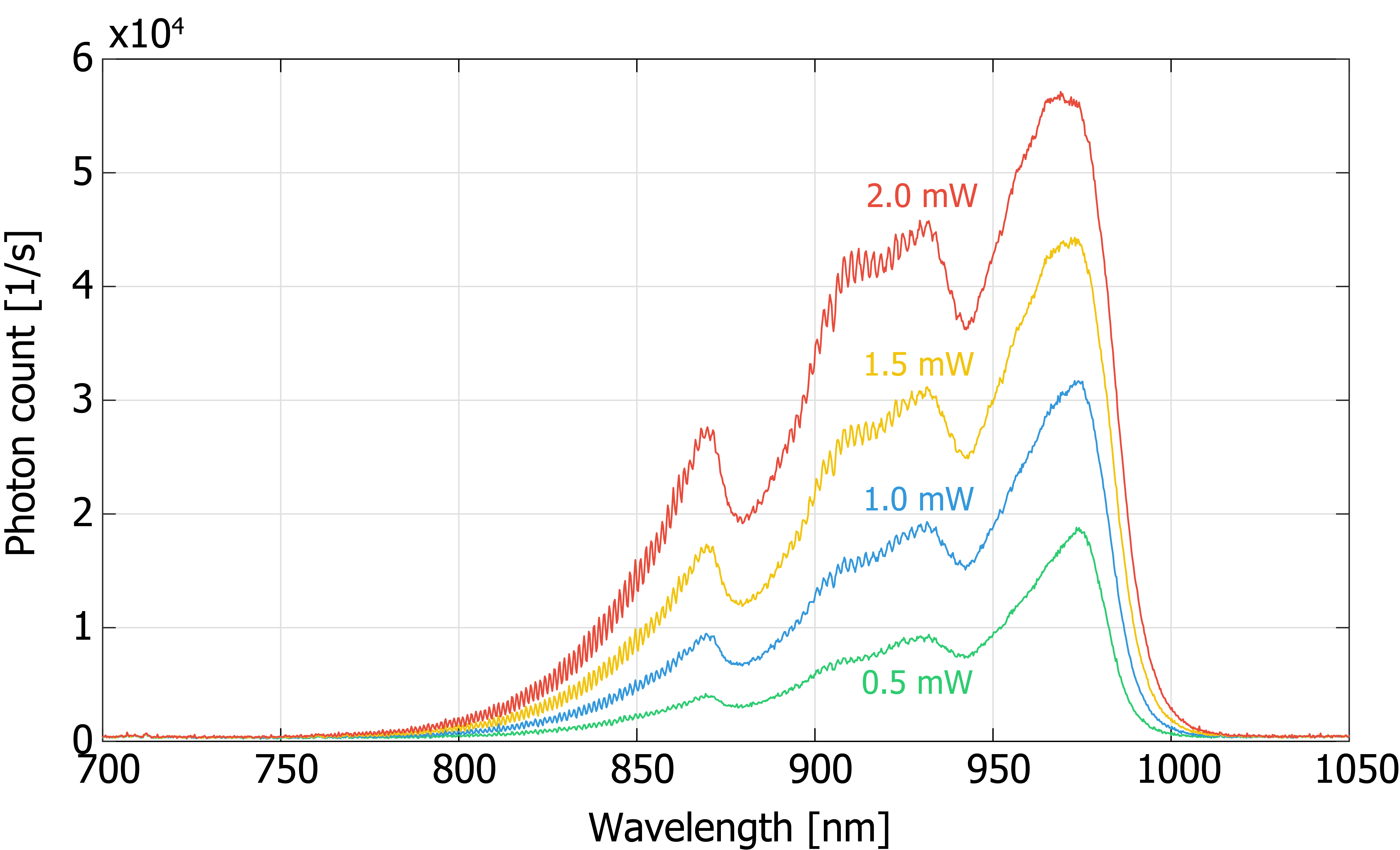}
\caption{\label{PL} Measured QW Photoluminescence (PL) at different pump powers. PL is pumped with a continuous wave laser at 637~\si{nm}.}
\end{figure}

Figure~\ref{PL} presents the measured photoluminescence (PL) spectra of the double QW active region. We pump the PL with a continuous wave (CW) laser operating at 637~\si{nm} at different powers. The laser light is focused on the top surface of the MBE-grown wafer with a 20x microscope objective. We then collect the generated PL with the same microscope objective, filter the pump light, and record the generated spectra with a grating spectrometer and CCD camera. We measure the peak PL emission wavelength of 972~\si{nm} at room temperature.

\section{76.2 mm diameter wafer-scale bonding}
\begin{figure}[ht]
\centering
\includegraphics[width=6cm]{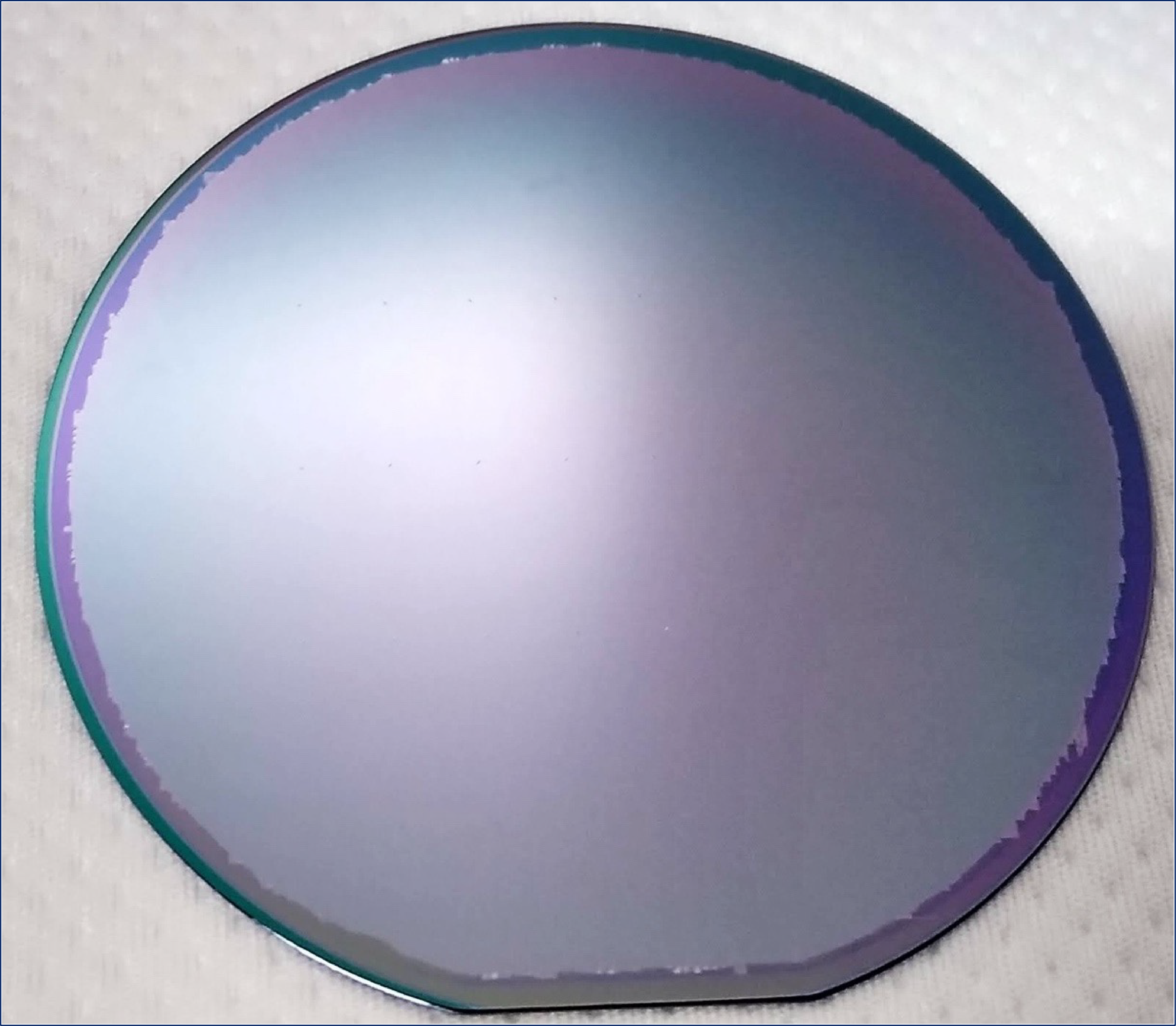}
\caption{\label{bonding} Camera image of the 76.2~\si{mm} diameter III-V gain material bonded to the tantala PIC wafer. Camera picture is taken after backside GaAs wafer removal and before patterning and etching of the active regions.}
\end{figure}

Figure~\ref{bonding} presents a camera image of the 76.2~mm diameter tantala PIC wafer after bonding the III-V gain material and backside GaAs wafer removal. The wafer-scale bonding process has a 3 mm wide wafer edge exclusion. We, however, measure \>~95~\% of the surface area yield.

\section{Multistage III-V-to-tantala transition taper}
\begin{figure}[htbp]
\centering
\includegraphics[width=\textwidth]{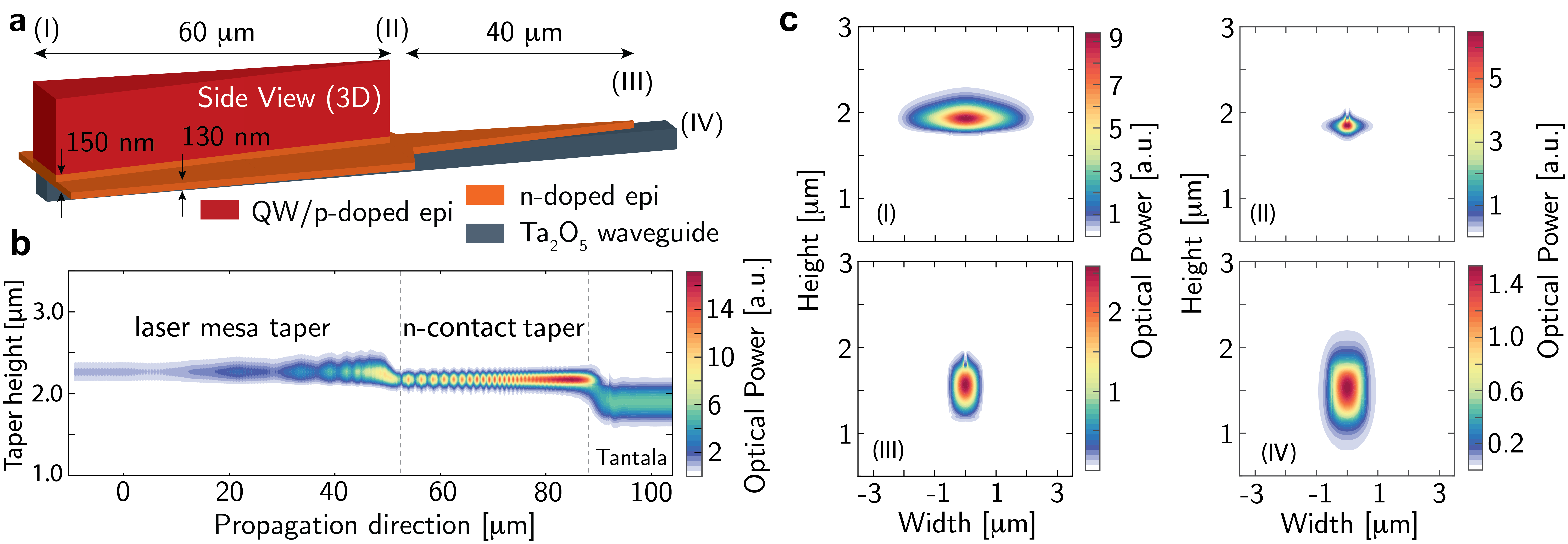}
\caption{\label{taper} III-V-to-tantala multistage transition taper. (a) Schematic diagram of the multistage taper presenting the three-dimensional (3D) perspective view of the laser mesa and n-contact tapers. (b) FDTD simulation of the optical mode propagation along the taper structure, facilitating the TE0 mode coupling from the laser mesa (on the left) to the tantala waveguide (on the right side). (c) Electric field intensity profiles (optical field profiles) of the TE0 mode at the input of the taper, the tip of the laser mesa taper, tip of the n-contact taper, and tantala waveguide presented in panels (I), (II), (III), and (IV), respectively.}
\end{figure}

A coupling structure is needed for efficient optical mode coupling from the III-V active region to the tantala photonic integrated circuit (PIC). We design the coupling structure based on multistage inverse tapers (Fig.~\ref{taper}a). First, the 1.2~\textmu{}m tall laser mesa is tapered from the nominal active region width to a 100~\si{nm} wide tip with the taper length of 60~\textmu{}m. The narrow taper tip structure with the 12:1 aspect ratio does not support optical mode confinement at 980~\si{nm} and the optical mode is coupled to the 150~\si{nm} thick n-contact layer. Next, we taper the n-contact layer from its nominal width of 100~\textmu{}m to 100~\si{nm} in two stages. In the first stage, we design and form a fast taper to reduce the n-contact width to 1.1~\textmu{}m. In the second stage, the width of the n-contact taper is further reduced to the final tip width of 100~\si{nm} using a 40~\textmu{}m long adiabatic taper structure to couple the optical mode to the 1~\textmu{}m wide tantala waveguide underneath the active region.

Figure~\ref{taper}b presents the Finite-Domain Time-Difference (FDTD) simulation of the fundamental-TE (TE0) optical mode propagation along the taper length, highlighting the mode transition from the laser mesa first to the n-contact layer and then to the tantala waveguide. Figure~\ref{taper}c presents the simulated optical mode profiles at different cross-sections along the length of the taper, highlighting the adiabatic nature of the taper. This enables the TE0 mode propagation without coupling to higher order modes supported by the structure. The simulated coupling efficiency of the structure is 86~\%.

\section{Distributed feedback laser wavelength control}
\begin{figure}[h!]
\centering
\includegraphics[width=6cm]{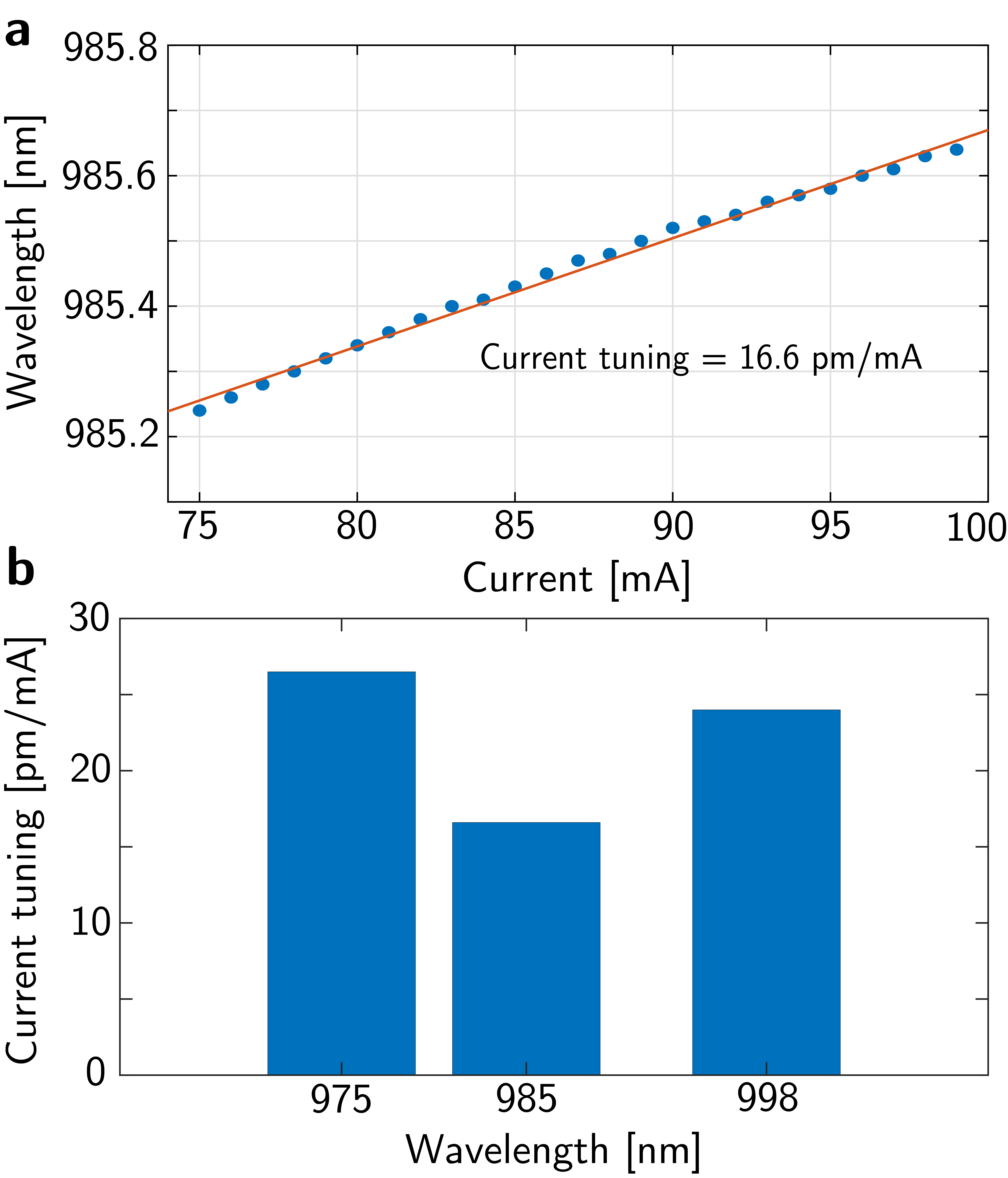}
\caption{\label{tuning} Injection-current-dependent wavelength-tuning of the DFB laser. (a) Lasing wavelength as a function of injection-current for a DFB laser operating at 985~\si{nm }. Measured data is presented with the scattered data points and the red line is the linear fit to the data with the fitted current-dependent tuning rate of 16.6 pm/mA. (b) Measured current-dependent tuning rates for three different DFB lasers operating at 975~\si{nm}, 985~\si{nm}, and 998~\si{nm}.}
\end{figure}

We measure single-mode injection-current-dependent wavelength tuning range of 300~GHz in the fabricated distributed feedback (DFB) lasers. This is due to resistive heating at different current levels. The dissipated electrical power, due to the series resistance of the laser, increases the active region temperature. This results in a redshift of the lasing wavelength with linear dependence with respect to the injection-current. Figure~\ref{tuning}a present the measured lasing wavelength as a function of the injection current with scattered data-points for a DFB laser operating at 985~\si{nm} with feedback grating period of 150~\si{nm}. The red line is the linear fit to the data with the calculated current-dependent tuning rate of 16.6 pm/mA. Figure~\ref{tuning}b summarizes the measured current tuning rates for three DFB lasers tested in the work. In addition to the 985~\si{nm} laser, we measure tuning rates of 26.6~pm/mA and 24.6 pm/mA for lasers operating at 975~\si{nm} and 998~\si{nm}, respectively.  

\section{Fabry-Perot lasers}
\begin{figure}[ht!]
\centering
\includegraphics[width=10cm]{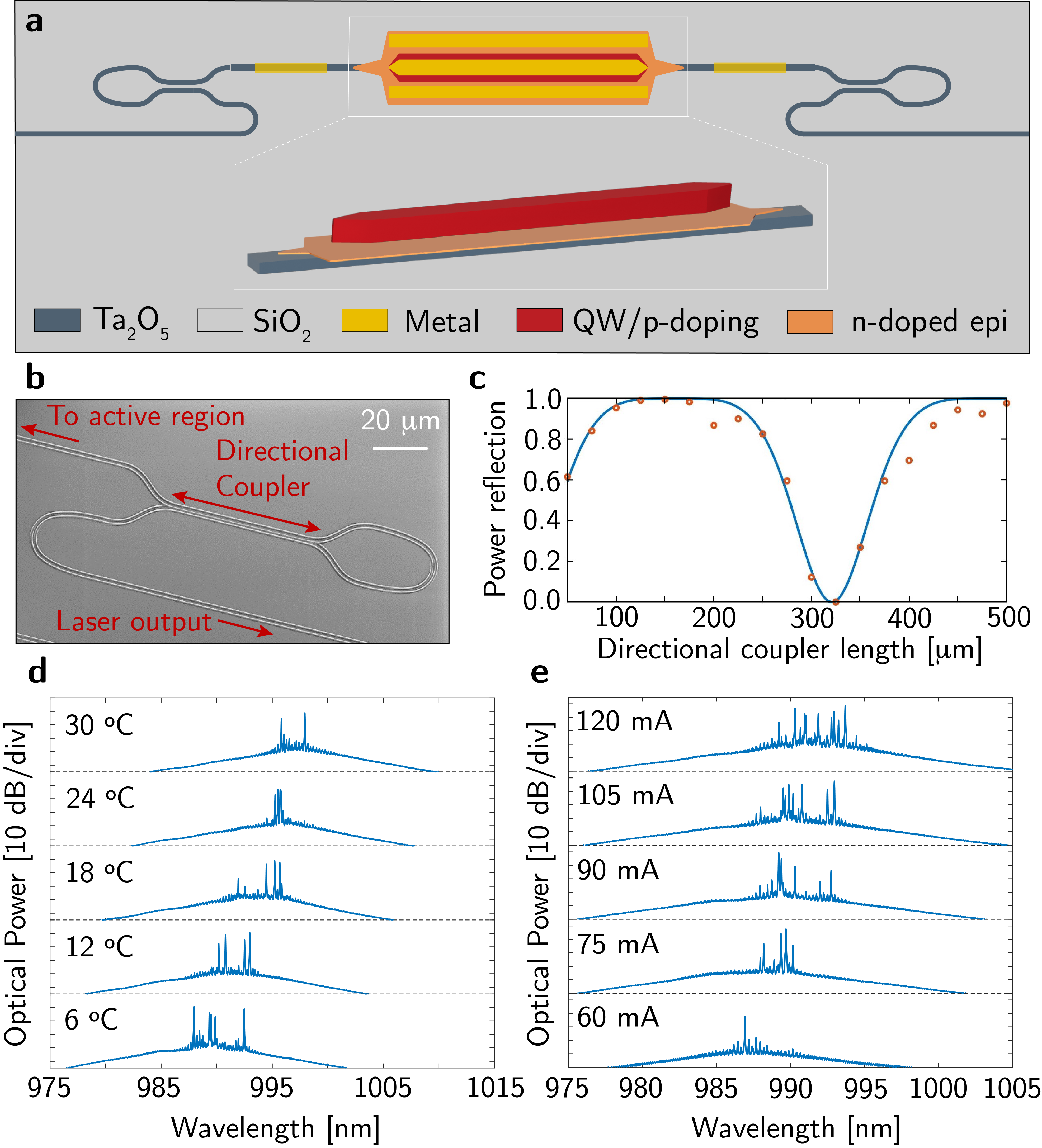}
\caption{\label{FP} (a) Schematic diagram of a Fabry-Perot laser with integrated loop-mirrors. (b) Scanning electron micrograph (SEM) of a fabricated loop-mirror. (c) Measured optical power reflection of a mirror as a function of the directional coupler length. (d) Stage-temperature and (e) injection-current tuning of a FP laser spectrum.}
\end{figure}

Figure~\ref{FP}a presents the schematic diagram of the fabricated Fabry-Perot (FP) lasers. We form the active region with a process and geometry similar to the DFB devices. We implement III-V-to-tantala transition tapers to couple the generated optical mode to loop-mirrors fabricated in the tantala waveguiding layer for cavity feedback (Fig.~\ref{FP}b). The loop-mirror consists of a directional coupler (DC) with its two branches connected on one side with a Sagnac loop structure. On the other side, one branch is connected to the active region, acting as the mirror input. The other branch is connected to the angled-facet edge couplers as the laser output port. The optical power reflection in the loop-mirror is based on optical mode coupling in the DC section, and the mirror structure does not contain any narrow-band, wavelength-selective structures. Therefore, a loop-mirror provides the broadband reflection needed for a FP laser with a power reflection level that is proportional to the DC length (Fig.~\ref{FP}c). 

Figures~\ref{FP}d and \ref{FP}e present the output spectra of a FP laser at different stage-temperatures and injection-currents, respectively. Data is plotted for the device whose voltage-current and power-current curves are plotted in the main text Fig.~4d. As expected, the generated multi-mode spectra redshifts with increasing stage-temperature and injection-current (due to the increasing resistive heating). The measured lasing modes are redshifted when compared to the measured room-temperature PL spectra in Fig.~\ref{PL}. This is true even for low stage-temperatures and injection-current. We attribute this redshift to localized resistive heating in the QW active region while the device is operating above threshold.

